\documentclass[twocolumn,epjc3]{svjour3}
\usepackage{graphicx}
\usepackage{amsmath}
\usepackage{amssymb}
\usepackage{bbm}
\usepackage{bm}
\usepackage{amsfonts}
\usepackage{epstopdf}
\usepackage[colorlinks,linkcolor=red,anchorcolor=blue,citecolor=green]{hyperref}
\usepackage{subfigure}
\usepackage{float}
\usepackage{enumerate}
\usepackage{cite}
\usepackage{fix-cm}
\newcommand\diff{\mathrm{d}}
\newcommand\Diff{\mathcal{D}}

\newcommand\e{\mathrm{e}}

\begin{document}
\title{
	Wormholes without exotic matter in nonminimal torsion-matter coupling $f(T)$ gravity
}
\author{Rui-Hui Lin\thanksref{e1,addr} \and Zhen-Yuan Wu\thanksref{addr} \and Xiang-Hua Zhai\thanksref{e2,addr}}
\thankstext{e1}{linrh@shnu.edu.cn}
\thankstext{e2}{zhaixh@shnu.edu.cn}
\institute{Shanghai United Center for Astrophysics (SUCA), Shanghai Normal University,
100 Guilin Road, Shanghai 200234, China\label{addr}}

\maketitle
\begin{abstract}
Wormholes are hypothetical tunnels that connect remote parts of spacetime.
In General Relativity, wormholes are threaded by exotic matter that violates the energy conditions.
In this work, we consider wormholes threaded by nonexotic matter in nonminimal torsion-matter coupling $f(T)$ gravity.
We find that the nonminimal torsion-matter coupling can indeed hold the wormhole open.
However, from geometric point of view, for the wormhole to have asymptotic flatness,
the coupling matter density must falloff rapidly at large radius,
otherwise the physical wormhole must be finite due to either change of metric signature
or lack of valid embedding.
On the other hand,
the matter source supporting the wormhole can satisfy the null energy condition only in the neighborhood of the throat of the wormhole.
Therefore, the wormhole in the underlying model has finite sizes and cannot stretch to the entire spacetime.
\end{abstract}
\section{Introduction}
Wormholes are theoretical solutions in General Relativity (GR) and other gravitation theories,
which describe geometrical structures that connect distant regions or different manifolds of spacetime.
Such a structure might be considered suitable for interstellar travel
or constructing closed timelike curve\cite{Morris1988,Visser1995}.
Nonetheless, in GR, holding the wormhole open needs exotic matter that violates the null energy conditions (NEC),
i.e. $\mathcal T_{\mu\nu}k^\mu k^\nu<0$,
where $\mathcal T_{\mu\nu}$ is the energy momentum tensor of the matter and $k^\mu$ is any null vector.
In fact, the wormhole geometry means a flaring out of the initially converging congruence of world lines,
which can be seen from the Raychaudhuri equation\cite{Visser1995}.
In the framework of GR,
this leads to the violation of energy conditions.
Therefore, in its weakest version, where the congruence of world lines is lightlike,
exotic matter that violates the NEC is required to support the wormhole.
This seems to be problematic in that
one may observe negative energy density while traveling through the wormhole.

Like its counterpart in cosmological scenarios,
the exotic content issue in wormhole configuration may also be investigated in terms of modified gravities.
The modifications of gravitation theory may provide extra terms that support the wormhole,
such that it is the effective energy momentum tensor $\mathcal T_{\mu\nu}^\text{eff}$ that violates the energy conditions,
while the matter threading the wormhole remains nonexotic and physical.
One of the simplest schemes of these modifications is the $f(R)$ gravity
\cite{Sotiriou2010,DeFelice2010,NOJIRI201159}.
Specifically, the higher order curvature terms in $f(R)$ gravity
\cite{Lobo2009,Pavlovic2015,Bambi2016,Mazharimousavi2016a,Mazharimousavi2016}
may help hold the wormhole open.
The wormhole geometries and energy conditions have also been studied in the further extensions of $f(R)$ gravity
including the nonminimal coupling $f(R,\mathcal L_\text M)$ gravity\cite{Garcia2010,Garcia2011,Harko2013}
where $\mathcal L_\text M$ is the Lagrangian of the matter,
and the $f(R,\mathcal T)$ gravity\cite{Zubair2016,Moraes2017,Sahoo2018}
where $\mathcal T$ is the trace of the physical energy momentum tensor.
In other modified gravities such as Einstein-Gauss-Bonnet theory\cite{Maeda2008,Kanti2011}, conformal gravity\cite{Hohmann2018a,Lobo2008},
scalar-tensor gravity\cite{Shaikh2016}
and Eddington-Born-Infeld gravity\cite{Shaikh2018,BELTRANJIMENEZ20181},
wormhole solutions have also been considered.

Besides the curvature-based alternatives of gravity,
one can also modify the gravitation theory starting from
the Teleparallel Equivalent of General Relativity (TEGR)\cite{aldrovandi2012teleparallel,Maluf2013}.
In the Lagrangian of TEGR,
the torsion scalar $T$ takes the place of the Ricci scalar $R$,
and hence, in analogous of $f(R)$,
$f(T)$ gravity\cite{Bengochea2009,Linder2010,Cai2016} and
its nonminimal coupling extension\cite{Harko2014a,Feng2015,Lin2017} have been studied.
Under the teleparallel framework,
the exotic content in wormholes has also been investigated
in $f(T)$ gravity\cite{Bohmer2012,Sharif2013} and its further extensions such as
nonminimal scalar-torsion coupling teleparallel gravity\cite{Bahamonde2016} and $f(T,T_G)$ gravity\cite{SHARIF2018145},
where $T_G$ is the teleparallel equivalent of the Gauss-Bonnet term.

In this work, we are concerned with the wormholes in the nonminimal torsion-matter coupling $f(T)$ gravity.
We consider the wormhole geometry and the NEC in various cases of a concrete model,
and find that the modification of gravitation involving coupling between torsion and matter
can indeed support the wormhole.
We illustrate the different qualitative behaviors of wormholes in these cases.
The paper is organized as follows.
In Sec. \ref{whintele} , we briefly review the nonminimal torsion-matter coupling $f(T)$ gravity
and the wormhole geometry.
We consider the wormhole solutions and the NEC with a concrete model in Sec. \ref{cases}.
Section \ref{conclusion} contains our main conclusions and discussions.
We will be using the unit $8\pi G=c=1$ throughout the paper.

\section{Wormhole geometry in nonminimal torsion-matter coupling gravity}
\label{whintele}
In this section, we briefly review the basics of the nonminimal torsion-matter coupling $f(T)$ gravity.
The geometry and the energy condition of the wormhole are also discussed.
\subsection{The nonminimal coupling $f(T)$ gravity}
\label{nonminimal}
On the spacetime manifold $\mathcal M$ with metric $g$
\begin{equation}
	g=g_{\mu\nu}\diff x^\mu\otimes\diff x^\nu,
	\label{metric}
\end{equation}
one can generally find dual pairs of linearly independent frame fields
or tetrad fields $\{e_a,e^a\}$,
such that $e^a\left( e_b \right)=\delta^a_b$ and
\begin{equation}
	\eta_{ab}=g\left( e_a,e_b \right).
	\label{etog0}
\end{equation}
In coordinate basis this can be expressed as
\begin{equation}
	\eta_{ab}=g_{\mu\nu}e_a^{\:\mu}e_b^{\:\nu}\quad\text{or}\quad g_{\mu\nu}=\eta_{ab}e^a_{\:\mu}e^b_{\:\nu}.
	\label{eg}
\end{equation}
And hence the determinant of the tetrad is
\begin{equation}
	|e|=\mathrm{det}(e^a_{\:\mu})=\sqrt{-g}.
	\label{edet}
\end{equation}
From Eq.\eqref{eg}, one can see that for a Lorentz rotation of the tetrad
\begin{equation}
	e_a^{\:\mu}\mapsto \tilde{e}_b^{\:\mu}=\Lambda_b^a e_a^{\:\mu},
	\label{lorentzrot}
\end{equation}
the metric remains the same.
And hence there could be infinite tetrads in a Lorentz group corresponding to a given metric.

The torsion tensor is given by\cite{aldrovandi2012teleparallel,Krssak2016,Golovnev2017}
\begin{equation}
	\begin{split}
	T^a_{\;\mu\nu}=&\Diff_\mu e^a_{\:\nu}-\Diff_\nu e^a_{\:\mu}\\
	=&\partial_\mu e^a_{\:\nu}+\omega^a_{\;b\nu}e^b_{\:\nu}-\partial_\nu e^a_{\:\mu}-\omega^a_{\;b\mu}e^b_{\:\mu},
\end{split}
	\label{Tpt}
\end{equation}
where the covariant derivative $\Diff_\mu$ and the spin connection $\omega^a_{\;b\mu}$ are introduced
such that for any vector $V^a$ in the tangent space, $\Diff_\mu V^a$ is locally Lorentz covariant under the rotation indicated by Eq.\eqref{lorentzrot}.
Among the infinite choices of tetrad,
it is generally possible to find a particular frame, the proper frame\cite{Lucas2009,Krssak2016,Lin2019},
in which all components of the spin connection vanish.
Our discussion then proceeds in such a frame and the spin connection terms are suppressed hereinafter.
With the torsion tensor one can construct the contorsion tensor
\begin{equation}
	K^{\mu\nu}_{\quad\rho}=-\frac12\left(T^{\mu\nu}_{\quad\rho}-T^{\nu\mu}_{\quad\rho}-T_{\rho}^{\:\mu\nu}\right)
	\label{contorsion}
\end{equation}
and the superpotential tensor
\begin{equation}
	S_\rho^{\:\mu\nu}=\frac12\left(K^{\mu\nu}_{\quad\rho}+\delta_\rho^\mu T^{\lambda\nu}_{\quad\lambda}-\delta_\rho^\nu T^{\lambda\mu}_{\quad\lambda}\right),
	\label{sptensor}
\end{equation}
where $T^\rho_{\:\mu\nu}=e_a^{\:\rho}T^a_{\:\mu\nu}$.
And then the torsion scalar is defined by
\begin{equation}
	T=T^\rho_{\:\mu\nu}S_\rho^{\:\mu\nu},
	\label{torsionscalar}
\end{equation}
which is used as the Lagrangian in TEGR\cite{aldrovandi2012teleparallel,Maluf2013}.
The nonminimal torsion matter coupling $f(T)$ gravity\cite{Harko2014a}
extends the usual $f(T)$ gravities\cite{Bengochea2009,Linder2010,Cai2016},
and adopts the following unifying form of action\cite{Feng2015,Lin2017,Lin2017jcap}
\begin{equation}
	\begin{split}
	\mathcal S=&-\frac12\int|e|(1+f_1(T))T\diff^4x\\
	&+\int|e|(1+f_2(T))\mathcal L_\text M\diff^4x.
\end{split}
	\label{ftaction}
\end{equation}
The field equation reads
\begin{equation}
	\begin{split}
		&\frac4{|e|}F\partial_\beta(|e|S_\sigma^{\:\:\alpha\beta}e_a^{\:\sigma})+4e_a^{\:\sigma}S_\sigma^{\:\:\alpha\beta}\partial_\beta F+4FS_\rho^{\:\:\alpha\sigma}T^\rho_{\:\:\sigma\beta}e_a^{\:\beta}\\
		&+(1+f_1)e_a^{\:\alpha}=2\mathcal{T}_\beta^{\:\alpha}e_a^{\:\beta},
	\end{split}
	\label{eom}
\end{equation}
where $F=1+f_1+f_{1T}-16\pi Gf_{2T}\mathcal L_\text M$,
$f_{iT}$ denotes the derivative of $f_i$ with respect to $T$
and $\mathcal T^\alpha_\beta$ is the energy momentum tensor of matter given by
\begin{equation}
	\frac{\delta(|e|\mathcal L_\text M)}{\delta e^a_{\:\alpha}}=|e|\mathcal T^\alpha_\beta e^a_{\:\beta}.
	\label{emtensor}
\end{equation}
Since the Lagrangian of matter $\mathcal L_\text M$ appears in the field equation \eqref{eom},
its explicit form is needed.
We assume the study of $\mathcal L_\text M$ for perfect fluid in the literature\cite{Minazzoli2012,Feng2015,Lin2017,Lin2017jcap} is applicable to the current work
and set $\mathcal L_\text M=\rho$ where $\rho$ is the energy density of matter.
\subsection{Wormhole geometry}
\label{traverwh}
As mentioned in the Introduction,
the wormholes in GR or TEGR\cite{Morris1988,Visser1995} require exotic matter that violates the NEC.
The investigation of wormholes threaded by nonexotic matter and supported by modifications of gravitation theory has been an active topic
\cite{Lobo2008,Maeda2008,Lobo2009,Garcia2010,Garcia2011,Kanti2011,Bohmer2012,Harko2013,Sharif2013,Pavlovic2015,Mazharimousavi2016,Mazharimousavi2016a,Bahamonde2016,Zubair2016,Bambi2016,Shaikh2016,Moraes2017,Hohmann2018a,Sahoo2018,Shaikh2018,SHARIF2018145} in recent years.

The notion of static spherically symmetric wormhole is concretized by the metric\cite{Morris1988}
\begin{equation}
	\diff s^2=\e^{2\varphi(r)}\diff t^2-\frac1{1-\frac{b(r)}r}\diff r^2-r^2\diff\Omega^2,
	\label{whmetric}
\end{equation}
where $\varphi(r)$ is the redshift function of radial coordinate $r$,
and $b(r)$ is the shape function.
To avoid horizon and singularity, the redshift function needs to be finite everywhere.
For simplicity, in this work we assume the redshift function as a constant,
and hence it could be absorbed into the temporal coordinate by a universal rescale of time.
The wormhole configuration requires that the radial coordinate $r$ have a nonzero minimum $r_0$,
i.e. the throat, at which the shape function has $b(r_0)=r_0$.
For an open wormhole, $b(r)$ also needs to satisfy the flaring-out condition\cite{Morris1988}
\begin{equation}
	\frac{b-b'r}{b^2}>0,
	\label{f-o}
\end{equation}
where $b'$ denotes the derivative of $b(r)$ with respect to $r$.

As discussed in Sec.\ref{nonminimal},
for a given metric, there can be infinite choices of corresponding tetrads related by Lorentz transforms.
Yet there is a particular one, the proper tetrad,
in which the spin connection vanishes.
For a static spherically symmetric metric like Eq.\eqref{whmetric},
a specific choice of proper tetrad has been proposed\cite{Lucas2009,Krssak2016}:
\begin{equation}
	e^a_{\:\mu}=\left(
	\begin{array}{cccc}
		1&0&0&0\\
		0&\frac{\sin\theta\cos\phi}{\sqrt{1-\frac br}}&r\cos\theta\cos\phi&-r\sin\theta\sin\phi\\
		0&\frac{\sin\theta\sin\phi}{\sqrt{1-\frac br}}&r\cos\theta\sin\phi&r\sin\theta\cos\phi\\
		0&\frac{\cos\theta}{\sqrt{1-\frac br}}&-r\sin\theta&0
	\end{array}
	\right).
	\label{propertetrad}
\end{equation}
We will continue our discussion using this proper tetrad.
The torsion scalar \eqref{torsionscalar} is then
\begin{equation}
	T(r)=\frac2{r^2}\left[ 2\left( 1-\sqrt{1-\frac{b(r)}r} \right)-\frac{b(r)}r \right].
	\label{tscalar}
\end{equation}

For anisotropic fluid the energy momentum tensor takes the form\cite{Morris1988}
\begin{equation}
	\mathcal T_{\mu\nu}=(\rho+p_t)u_\mu u_\nu-p_tg_{\mu\nu}+(p_r-p_t)\chi_\mu\chi_\nu,
	\label{anisoemtensor}
\end{equation}
where $\rho,p_r$ and $p_t$ are the energy density, radial pressure and tangential pressure of matter, respectively.
And $u_\mu$ and $\chi_\mu$ are the 4-velocity and the unit spacelike 4-vector, satisfying $u^\mu u_\mu=1$, $\chi^\mu\chi_\mu=-1$ and $u^\mu\chi_\mu=0$.
With Eq.\eqref{anisoemtensor},
$\mathcal T_{\mu\nu}u^\mu u^\nu\ge0$ leads to $\rho\ge0$,
and the NEC $\mathcal T_{\mu\nu}k^\mu k^\nu\ge0$ leads to $\rho+p_r\ge0$.

The components of the field equation \eqref{eom} can be written as
	\begin{equation}
		\begin{split}
			\frac{2Fb'}{r^2}=&2(1+f_2)\rho+(F-f_1-1)T\\
			     &-\frac4r\sqrt{1-\frac br}\left( 1-\sqrt{1-\frac br}F' \right),\\
			\frac{2Fb}{r^3}=&-2(1+f_2)p_r+(F-f_1-1)T,\\
			\frac{F(b'r-b)}{r^3}=&-2(1+f_2)p_t+(F-f_1-1)T\\
				 &-\frac2r\sqrt{1-\frac br}\left( 1-\sqrt{1-\frac br}F' \right),
		\end{split}
		\label{eom2}
	\end{equation}
where the prime indicates the derivative with respect to the radial coordinate $r$.

\section{Wormhole solutions in a concrete model}
\label{cases}
The field equation \eqref{eom2} describes the relation between the matter source $\rho,p_r,p_t$ and the wormhole geometry $b(r)$.
For the system to be determinative,
in this work we adopt the monotonically decreasing energy density profile\cite{Garcia2010,Garcia2011}:
\begin{equation}
	\rho(r)=\rho_0\left( \frac{r_0}r \right)^\alpha,
	\label{densityprofile}
\end{equation}
where $\alpha$ is a constant with $\alpha>0$
and $\rho_0$ is the energy density at the throat $r_0$.
As a simple but heuristic model,
we also specify that 
\begin{equation}
	f_1=0,\quad\text{and}\quad f_2=\lambda T,
	\label{model}
\end{equation}
where $\lambda$ is the parameter of the model with the dimension of square of length.

Then the field equation \eqref{eom2} leads to
\begin{equation}
	b'=\frac{r^2}{1-2\lambda\rho}\left[ \rho+\frac4r\lambda\rho'\sqrt{1-\frac br}\left( 1-\sqrt{1-\frac br} \right) \right],
	\label{b1}
\end{equation}
and
\begin{equation}
	p_r=-\frac1{1+\lambda T}\left[ \lambda\rho T+\frac b{r^3}\left( 1-2\lambda\rho \right) \right].
	\label{pr1}
\end{equation}
For a typical wormhole solution,
the shape function $b(r)$ needs to satisfy the flaring-out condition \eqref{f-o} in the neighborhood of the throat $r_0$, i.e. $b'|_{r_0}<1$,
which leads to constraints of the model parameter $\lambda$
\begin{equation}
	\lambda<\frac{1-r_0^2\rho_0}{2\rho_0},\quad\text{or}\quad\lambda>\frac1{2\rho_0}.
	\label{lambda11}
\end{equation}
Moreover, we wish to find wormholes threaded by nonexotic matter that satisfies the NEC at least at the throat $r_0$, i.e. $\rho(r_0)+p_r(r_0)>0$.
With Eqs.\eqref{densityprofile} and \eqref{pr1}, we have
\begin{equation}
	\rho(r_0)+p_r(r_0)=\rho_0-\frac1{(2\lambda+r_0^2)}>0.
	\label{nec1}
\end{equation}
This gives
\begin{equation}
	\lambda<-\frac{r_0^2}2,\quad\text{or}\quad\lambda>\frac{1-\rho_0r_0^2}{2\rho_0}.
	\label{lambda12}
\end{equation}
The interception of the inequalities \eqref{lambda11} and \eqref{lambda12} then gives the combined constraint on $\lambda$
\begin{equation}
	\lambda>\frac1{2\rho_0},\quad\text{or}\quad\lambda<-\frac{r_0^2}2.
	\label{lambda1}
\end{equation}

\subsection{Asymptotically flat case}
\begin{figure}[!ht]
	\centering
	\includegraphics[width=0.9\linewidth]{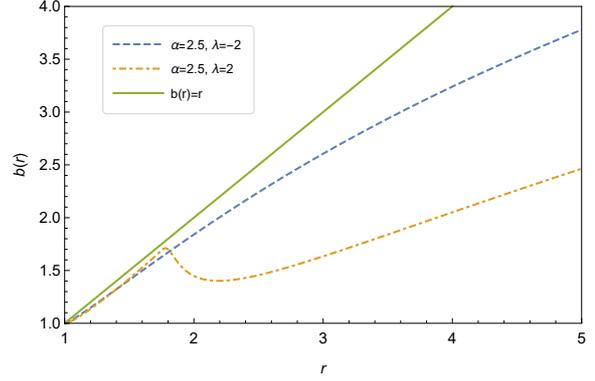}
	\caption{The numerical solution of $b(r)$ with $\alpha=2.5$ and $\lambda=\pm2$,
	where we have set $\rho_0=1$ and $r_0=1$.}
	\label{m1flatb}
\end{figure}

Additionally, one may consider the wormhole spacetime to be asymptotically flat,
which means
\begin{equation}
	\frac{b(r)}r\rightarrow0,\quad b(r)>0,
	\label{asymflat}
\end{equation}
as $r\rightarrow\infty$.
The asymptotically behavior of Eq.\eqref{b1} at $r\rightarrow\infty$ is
\begin{equation}
	b'\simeq\frac{\rho_0r^2r_0^\alpha}{r^\alpha-2\lambda\rho_0r_0^\alpha}\left( 1-2\alpha\lambda \frac b{r^{3}} \right).
	\label{b1asym}
\end{equation}
This equation can be solved analytically as
\begin{equation}
	b(r)\simeq\left\{\begin{aligned}
		&\frac{c_1(\alpha-3)r^\alpha-\rho_0r_0^\alpha r^3}{(\alpha-3)(r^\alpha-2\rho_0r_0^\alpha\lambda)}&\text{if}\:\alpha\ne3,\\
		&\frac{r^3\left(c_1+\rho_0r_0^\alpha\log\left( \frac r{r_0} \right)\right)}{r^3-2\rho_0r_0^\alpha\lambda}&\text{if}\:\alpha=3,
	\end{aligned}\right.
	\label{b1asymsol}
\end{equation}
where $c_1$ is some integral constant.
The asymptotic flatness \eqref{asymflat} then requires that $\alpha>2$,
and that $c_1>0$ if $\alpha\ge3$.
Fig.\ref{m1flatb} shows the qualitative behaviors of the shape functions $b(r)$
for $\alpha>2$ and $\lambda$ in the ranges given by inequality \eqref{lambda1}.

\begin{figure}[!ht]
	\centering
	\includegraphics[width=0.9\linewidth]{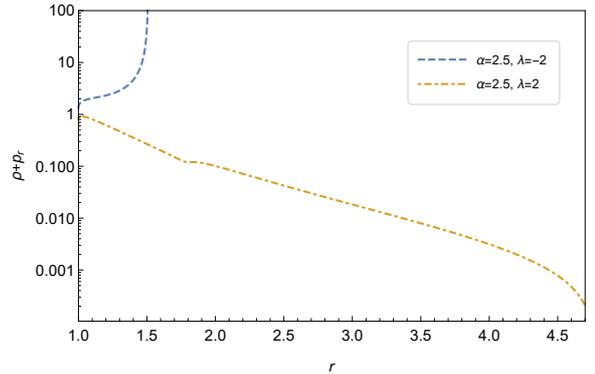}
	\caption{$\rho+p_r$ with $\alpha=2.5$ and $\lambda=\pm2$,
		where we have set $\rho_0=1$ and $r_0=1$.
In the case $\lambda=-2$, $\rho+p_r$ has a pole located at $r=1.51$;
In the case $\lambda=2$, $\rho+p_r$ will change to negative at $r=4.79$.}
	\label{m1flatpr}
\end{figure}

Though the NEC is ensured by inequality \eqref{lambda1} at the throat,
it still needs to be checked for the entire wormhole spacetime.
If asymptotic flatness Eq.\eqref{asymflat} is satisfied,
i.e. $\alpha>2$,
then $T\simeq\frac{b^2}{2r^4}$ as $r\rightarrow\infty$, and Eq.\eqref{pr1} becomes
\begin{equation}
	p_r=\left\{\begin{aligned}
		&-\frac{\rho_0r_0^\alpha}{3-\alpha}r^{-\alpha}+\mathcal{O}\left( r^{2-3\alpha} \right),&\text{if}\;2<\alpha<3,\\
		&-c_1r^{-3}+\mathcal{O}\left( r^{-7} \right),&\text{if}\;\alpha\ge3.
	\end{aligned}\right.
	\label{asympr1}
\end{equation}
Thus as $r\rightarrow\infty$,
\begin{equation}
	\rho+p_r=\left\{\begin{aligned}
		&\frac{2-\alpha}{3-\alpha}\rho_0r_0^\alpha\:r^{-\alpha}+\mathcal{O}\left( r^{2-3\alpha} \right),&\text{if}\;2<\alpha<3,\\
		&-c_1r^{-3}+\mathcal{O}\left( r^{-\alpha} \right),&\text{if}\;\alpha\ge3.
	\end{aligned}\right.
	\label{asymnec1}
\end{equation}
This means that the NEC is always violated at large $r$.
Furthermore, if $\lambda<-r_0^2/2<0$,
the denominator of Eq.\eqref{pr1} will change sign as $r$ increases
and hence $\rho(r)+p_r(r)$ will encounter a pole.
Therefore, the wormholes in these cases cannot stretch to the entire spacetime,
and only the vicinity of the throat is physical, where $\rho(r)+p_r(r)$ is positive and finite, as plotted in Fig.\ref{m1flatpr}.
Outside this physical region,
the wormhole should be joined with a physical vacuum\cite{Garcia2011,Rosa2018}.

\subsection{Asymptotically non-flat case}
\begin{figure}[!ht]
	\centering
	\includegraphics[width=0.9\linewidth]{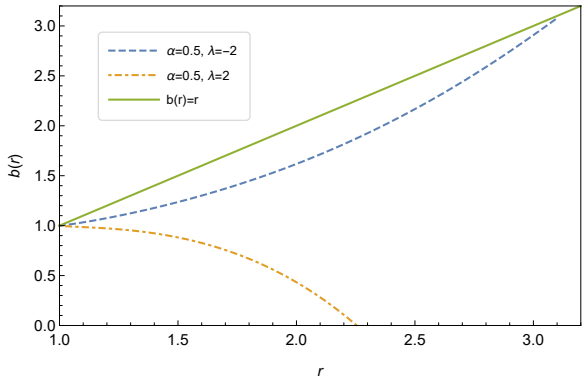}
	\caption{The numerical solution of $b(r)$ with $\alpha=0.5$ and $\lambda=\pm2$,
where we have set $\rho_0=1$ and $r_0=1$.
For $\lambda=-2$, $b(r)$ reaches
the solid line $b(r)=r$ at $r=3.13$.
For $\lambda=2$, $b(r)$ reaches zero at $r=2.25$.}
	\label{m1nonflatb}
\end{figure}

\begin{figure}[!ht]
	\centering
	\includegraphics[width=0.9\linewidth]{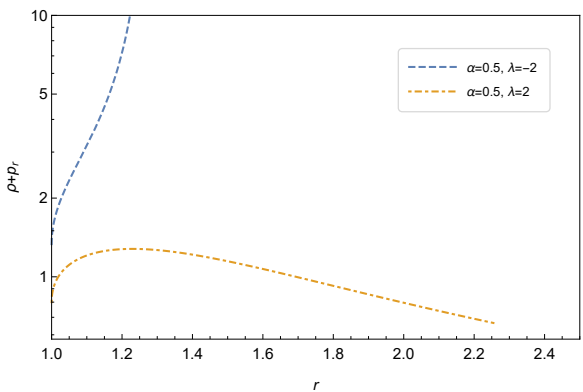}
	\caption{$\rho+p_r$ with $\alpha=0.5$ and $\lambda=\pm2$,
where we have set $\rho_0=1$ and $r_0=1$.
For $\lambda=-2$, $\rho+p_r$ encounters a pole at $r=1.28$.
}
	\label{m1nonflatpr}
\end{figure}

Alternatively,
one may loosen the asymptotically flat condition \eqref{asymflat}
and consider the case with $\alpha\le2$.
Fig.\ref{m1nonflatb} shows the qualitative behaviors of the shape functions $b(r)$
for $\alpha\le2$ and $\lambda$ in the ranges given by inequality \eqref{lambda1}.
And the NEC is checked in Fig.\ref{m1nonflatpr}.

For $\alpha<2$ and $\lambda>\frac{r_0^2}2$,
the shape function $b(r)$ will change from positive to negative as $r$ increases.
Within the region where $b(r)$ is positive, the NEC is satisfied, as showed in Fig.\ref{m1nonflatpr}.
Outside this region, the spacelike 3-geometry cannot be embedded in a four dimensional Euclidean space,
but in a Minkowski space\cite{Cataldo2008}.
Therefore, the spacetime in this case is no longer a physical wormhole outside this region.

For $\alpha<2$ and $\lambda<-\frac1{2\rho_0}$,
the shape function $b(r)$ will reach $b(r)=r$ as $r$ increases.
When $b(r)>r$, the $g_{rr}$ component of the metric Eq.\eqref{whmetric} changes sign, and the radial coordinate becomes timelike.
Moreover, $\rho(r)+p_r(r)$ will also encounter a pole as $r$ increases in this case.
Therefore, only the neighborhood where $\rho+p_r$ is finite and $b(r)<r$ is physical for wormhole configuration.

Nonetheless, one may also join the wormholes in these two cases with a physical vacuum outside the vicinity of the throat,
and make sure that the entire spacetime is physical.

\section{Conclusion and discussions}
\label{conclusion}
In this paper, we have considered the configuration of wormhole
in nonminimal torsion-matter coupling $f(T)$ gravity.
We focus on the wormhole geometry and the NEC for the threading matter in a simple but heuristic model: $f_1=0,f_2\propto T$.
Our main conclusions are as follows:
\begin{itemize}
	\item We find that the model can form wormholes without exotic matter that violates the NEC.
		Near the throats, the modification of TEGR involving nonminimal torsion-matter coupling can indeed provide support for the wormholes.
	\item However, the wormhole formed in this model cannot stretch to the entire spacetime and a physical vacuum should be connected to the wormhole:
		\begin{itemize}
			\item For the shape of the wormhole to be asymptotically flat,
				the matter needs to falloff rapidly as $r$ increases,
				as indicated by the requirement $\alpha>2$.
				As for the NEC, $\rho+p_r$ will be positive and finite only in some neighborhood of the throat.
			\item If $\alpha\le2$, at large radius, either the spacetime metric will change sign or the wormhole cannot have a valid embedding in Euclidean space.
				Within the vicinity of the throat,
				the wormhole can be held open by the nonminimal coupling to the matter,
				and the NEC can be satisfied.
		\end{itemize}
\end{itemize}

The flaring-out condition and the NEC only guarentee that the wormholes are open and threaded by nonexotic matter.
For the wormholes to be traversable,
there are other conditions such as traveling time and tidal forces that should be considered\cite{Morris1988},
which is beyond the scope of this work.
The traversability of the wormhole in nonminimal torsion-matter coupling $f(T)$ gravity will be considered in the forthcoming work.

Since all the wormhole configurations in this work have finite sizes in spacetime,
it is natural to consider the junction conditions at the edges of the wormholes connecting to the outside vacuum.
However, the Schwarzschild vacuum is not the solution to $f(T)$ gravities if the tetrad \eqref{propertetrad} is used as the proper tetrad\cite{Bohmer2011,Tamanini2012},
and the actual vacuum solutions in these cases are yet to be found.
On the other hand, if one chooses different proper tetrads\cite{Ferraro2011a,Lin2019},
the Schwarzschild or Shwarzschild-de Sitter solution may be admitted as a valid and physcial vacuum in $f(T)$ gravities.
This issue also worth considering in the future.

\bibliography{ref}
\bibliographystyle{spphys}

\end{document}